\documentclass[prb,twocolumn,showpacs,preprintnumbers,amsmath,amssymb]{revtex4}
\usepackage{graphicx}
\usepackage{dcolumn}
\usepackage{bm}
\begin{document}

\title{Two Anderson Impurities in the Kondo Limit\\ Doublet States}
\author{J. Simonin}
\affiliation{Centro At\'{o}mico Bariloche and Instituto Balseiro, \\
8400 S.C. de Bariloche, R\'{i}o Negro,  Argentina}
\date{mar}

\begin{abstract}
We analyze two pairs of doublet states for the two Anderson impurity
problem. We found that for short interimpurity distances they
have a lower energy than the ferro triplet and the antiferro
singlet. For medium to long distances between the impurities, the doublets
also have a lower energy than the two decoupled Kondo-singlet
state for most of the range of interest. The mechanism behind
their energy gain is an coherence enhanced Kondo-like
interaction.
\end{abstract}
\maketitle

The two Anderson impurity problem, i.e. how two magnetic impurities interact when they are embedded in a metallic host, has been the subject of many theoretical studies, ranging from perturbation theory \cite{coq}, \cite{cesar}, \cite{carlos} and narrow band approximation \cite{robert} to renormalization group analysis \cite{jones}. Besides its natural presence in alloys, today this system can be tailor made by means of quantum dots \cite{craig}, \cite{pascal} and promises to be a relevant circuit component in quantum electronics. Here we present a variational wave function analysis that, as in the single impurity case\cite{varma}, gives a clear description of the system. When needed, we specialize in the one dimensional (1D) case because of its technological applications.

The Anderson Hamiltonian for magnetic impurities diluted
in a metallic host is:
\begin{eqnarray}\label{hamil}
H=\sum_{k \sigma} e_k c^\dag_{k\sigma} c_{k\sigma} +
\frac{V}{\sqrt{N_c}}\sum_{j k \sigma}(e^{i\bf{k}.\bf{r_j}} \ \
d^\dag_{j\sigma} c_{k\sigma}+ h.c. )\nonumber \\
 - E_d \sum_{j \sigma} d^\dag_{j\sigma} d_{j\sigma} +
 U \sum_j d^\dag_{j\uparrow} d_{j\uparrow}d^\dag_{j\downarrow}d_{j\downarrow} \ \ ,
\end{eqnarray}
where the fermion operators $c_{k\sigma}$ act on the conduction
band states and $d_{j\sigma}$ on the orbital of the magnetic
impurity situated at $\bf{r_j}$. Single state energies $e_k,
-E_d$ are refereed to the Fermi energy ($E_F = k_F^2/2 m$), i.e.
there is an implicit $-\mu \ N$ term in the Hamiltonian that
regulates the population of the system ($\mu=E_F$ is the chemical
potential and $N$ the total number operator), $V$ is the $d-c$
hybridization and $N_c$ is the number of cells in the metal. In
the Kondo limit the impurity levels are well below the Fermi
energy ($E_d \gg 0$), and they can not be double occupied due to
the Coulomb repulsion in them ($U \gg 2 E_d$). In order to
simplify calculations we renormalize the vacuum (denoted by
$|F\rangle$) to be the conduction band filled up to the Fermi
energy and we make an electron-hole transformation for band states
below the Fermi level: $b^\dag_{k\sigma}\equiv c_{-k,-\sigma}$ for
$|k|\leq k_F$. In this way the energy of a hole excitation (i.e., to
remove an electron from below the Fermi level) is explicitly
positive.

We consider here the two impurities case, one impurity at $-
\textbf{r}/2$ (the Left impurity) and the other at $\textbf{r}/2$
(the Right impurity). Although calculations were made with the
second quantization operators, we use in the text a ``ket"
notation for the impurity states: the first symbol indicates the population status of the Left impurity and the second the status of the
Right impurity, e.g. $|\downarrow,\uparrow\rangle \equiv
 d^\dag_{L\downarrow} d^\dag_{R\uparrow} |F\rangle$,
$\ |0,\uparrow\rangle \equiv d^\dag_{R\uparrow}|F\rangle$,
 $\ |0,0\rangle \equiv|F\rangle$, etc..

In this situation ($E_d \gg 0$, $U \gg 2 E_d$) the lower energy
configurations of the system are the ferro triplet
($|\uparrow,\uparrow\rangle$, $\ (|\uparrow,\downarrow\rangle +
|\downarrow,\uparrow\rangle)$, $\ |\downarrow,\downarrow\rangle$)
and the antiferro singlet ($|\uparrow,\downarrow\rangle -
|\downarrow,\uparrow\rangle$), with energies $-2 E_d$. These
states are not eigenstates of the system; the hybridization mixes
these states with configurations having excitations in the band.
Perturbation theory allows for the calculation of the
corresponding corrections. These energy corrections are the same
for both the ferro (FF) and antiferro (AF) states (and twice the
single impurity corrections) up to the well known RKKY-like four
order term, which has opposite contributions on
them\cite{coq}\cite{cesar}:
\begin{eqnarray}\label{rkky1}
E_{FF}=-2E_d - \Sigma_R\ , \ \ \   E_{AF}=-2E_d + \Sigma_R\ ,    \\
\Sigma_R=2 \ \textbf{v}^4 \sum_{k,q}
\frac{\cos{((\bf{q}+\bf{k}).\bf{r})}}{(E_d+e_k)^2(e_k+e_q)}\ ,\ \ \ \ \ \ 
\end{eqnarray}
where the $q$ ($k$) sum is over hole (electron) excitations
($|q|\leq k_F$, $|k|\geq k_F$) and $\textbf{v}=V/\sqrt{N_c}$ .

A related state is the Kondo singlet, the ground state when just
one impurity is considered. If only the Left impurity is present,
this singlet is described by the variational wave
function\cite{varma}
\begin{equation}\label{wvarma}
|S_L\rangle=|F\rangle - \sum_q \alpha^K_q e^{i\textbf{q.r}/2}
(b^\dag_{q\uparrow}|\downarrow_L\rangle +
b^\dag_{q\downarrow}|\uparrow_L\rangle )\ ,
\end{equation}
where the variational parameter results to be
$\alpha^K_q=\textbf{v}/(E_S + E_d - e_q)$
 and a self-consistent equation is obtained for the singlet energy $E_S$,
\begin{equation}\label{es}
E_S=2\ \textbf{v}^2 \sum_{q}\frac{1}{E_S + E_d - e_q}\ ,
\end{equation}
after the pole in the sum, $E_S = -E_d - \delta_K$ is proposed
($E_d \gg \delta_K$), and Eq.(\ref{es}) becomes an equation for
the Kondo energy $\delta_K$,
\begin{equation}
E_d=2\ \textbf{v}^2 \sum_{q}\frac{1}{\delta_K + e_q}\ ,
\end{equation}
from which
\begin{equation}\label{ekondo}
\delta_K = D\ e^{-1/(2 Jn)}
\end{equation}
is obtained. $Jn$ ($\equiv(V^2/E_d)\ n_o$, $n_o$ being the density
of states at the Fermi level) is the relevant parameter for these
theories, and $D$, the half band-width, becomes the energy scale
($\delta_K$, $\Sigma_R$ are proportional to it). $Jn=0.07238$ gives $\delta_K$ one
thousandth of $D$, a reasonable value.

We study two doublet states, the even doublet,
\begin{eqnarray}\label{weven}
|D_E\rangle = (|\uparrow,0\rangle + |0,\uparrow\rangle)
+ \sum_{q}\ \{ \ \ \ \ \ \ \ \ \ \ \ \ \nonumber \\
-i \ \alpha^F_q \ \sin{\frac{q.r}{2}} \ [\ 2b^\dag_{q \downarrow}
|\uparrow,\uparrow\rangle + \ b^\dag_{q \uparrow}
(|\uparrow,\downarrow\rangle + |\downarrow,\uparrow\rangle)]\nonumber \\
 + \ \alpha^A_q \ \cos{\frac{q.r}{2}} \ \  b^\dag_{q \uparrow}\
(|\uparrow,\downarrow\rangle-|\downarrow,\uparrow\rangle)\ \}\ ,
\end{eqnarray}
and the odd doublet,
\begin{eqnarray}\label{wodd}
|D_O\rangle= (|\uparrow,0\rangle - |0,\uparrow\rangle)
+ \sum_{q} \ \{ \ \ \ \ \ \ \ \ \ \ \ \ \nonumber \\
+ \ \alpha^F_q \ \cos{\frac{q.r}{2}} \ [\ 2b^\dag_{q \downarrow}
|\uparrow,\uparrow\rangle + \ b^\dag_{q \uparrow}
(|\uparrow,\downarrow\rangle + |\downarrow,\uparrow\rangle)]\nonumber \\
 -i\ \alpha^A_q \ \sin{\frac{q.r}{2}} \ \  b^\dag_{q \uparrow}\
(|\uparrow,\downarrow\rangle-|\downarrow,\uparrow\rangle)\ \}\ .
\end{eqnarray}

After a standard minimization procedure, the variational
parameters are obtained
\begin{equation}\label{alfad}
\alpha^{F(A)}_q=\frac{\textbf{v}}{E_X+2E_d\pm\Sigma_R-e_q}\ ,
\end{equation}
where $E_X$ stands for the energy of the doublet ($E_E$ or $E_O$)
and the plus sign in front of $\Sigma_R$ holds for the ferro
$\alpha$ factor ($\alpha^F$) and the minus for the antiferro. For
the energies, self-consistent equations result
\begin{eqnarray}\label{enede}
E_E=-E_d+\textbf{v}^2\sum_{q}\frac{3\sin^2{\frac{q.r}{2}}}
{E_E+2E_d+\Sigma_R-e_q}\nonumber \\
+\textbf{v}^2\sum_{q}\frac{\cos^2{\frac{q.r}{2}}}
{E_E+2E_d-\Sigma_R-e_q}\ .
\end{eqnarray}
For $E_O$ a similar equation holds, with the $\sin$ and $\cos$
interchanged. This behavior, as well as the sign of $\Sigma_R$, is
due to the spatial symmetry of the ferro and antiferro states. As
is the case of the Kondo singlet, the dominant pole in the sums
of Eq.(\ref{enede}) determines the value of $E_E$.

First, let us analyze the situation when $\Sigma_R$ vanishes. This
situation corresponds to the zeros of the RKKY interaction and the
$k_F r \gg \pi$ limit.

For $\Sigma_R=0$, after expanding the $\sin^2$, $\cos^2$
functions, Eq.(\ref{enede}) reads
\begin{equation}\label{es01}
E_E=-E_d+\textbf{v}^2\sum_{q}\frac{2-\cos{q.r}} {E_E+2E_d-e_q}\ .
\end{equation}
Proposing  $E_E = -2 E_d - \delta_E$ we obtain
\begin{equation}\label{es02}
E_d=\textbf{v}^2\sum_{q}\frac{2-\cos{q.r}} {\delta_E+e_q}\ .
\end{equation}
Therefore for very large $r$, such that the contribution from the
$\cos$ term vanishes, the energy gain of the doublet ($\delta_E$)
tends to that of one Kondo singlet. This is not surprising, given
that the doublets, when $\Sigma_R=0$, can be written as
\begin{equation}
|D_X\rangle= |\uparrow_L\rangle \otimes|S_R\rangle \pm |S_L\rangle
\otimes|\uparrow_R\rangle,
\end{equation}
i.e., the combination of a Kondo singlet in one impurity and the
other single occupied plus (minus) their mirror image.

This energy, $\delta_E$, must be compared with twice the Kondo
energy, because, for very large $r$, the latter is the energy gain for a
state that has simultaneously both impurities forming a Kondo
singlet ($|S_L\rangle \otimes|S_R\rangle$). 

\begin{figure}[h]
\includegraphics[width=\columnwidth]{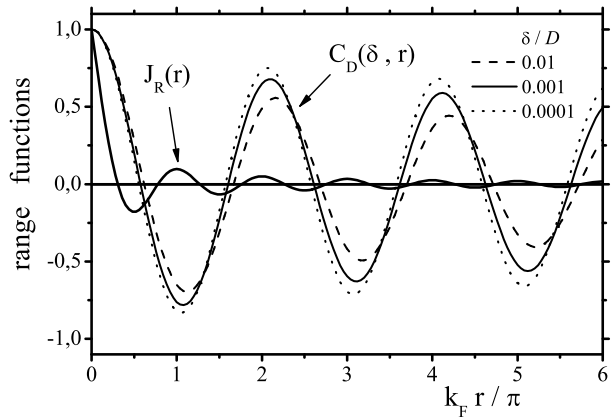}
\caption{Range functions for the RKKY, $J_R(2 k_F r)$, and the
Doublet Coherence factor, $C_D(\delta ,k_F r)$. The period of the RKKY is
half the one of $C_D$ and its amplitude decays very rapidly. These
characteristics are due to the fact that the RKKY depends on two
excitations, whereas the Doublet Coherence involves just one.}
\label{fig1}
\end{figure}

Now, we examine the effect of the coherence term in
Eq.(\ref{es02}); here we present the 1D analysis. The sums in Eq.(\ref{es02}) can be done exactly. The first one is the Kondo integral and we call the second one the Doublet Coherence integral,
\begin{eqnarray}
\frac{1}{N_c}\sum_{q}\frac{1}{\delta+e_q} &=&n_o \
I_K(\delta)=n_o \ \ln{(1+\frac{D}{\delta})}\ ,  \\
\frac{1}{N_c}\sum_{q}\frac{\cos{q.r}} {\delta+e_q}&=&n_o \
I_X(\delta, x)\ ,
\end{eqnarray}
where $x=k_F r$. The Coherence integral has a logarithmic
dependence on $\delta$ that goes like $I_K$, so it is useful to
define $C_D(\delta,x)=I_X(\delta,x)/I_K(\delta)$, which is a
decaying oscillatory function that depends lightly on $\delta$. We
will see that $C_D$ plays the role of a coherence driven extra
connectivity between the components of the doublet wave function.
In Fig.\ref{fig1} we plot $C_D$ as a function of $k_F r$ for
various values of $\delta$, and the $r$ dependence of the 1D-RKKY is
also plotted. It can be seen that the RKKY reaches its asymptotic
form ( $-\cos{(2x)}/\pi x$ ) very quickly, at $k_F r\geq \pi$,
whereas $I_X$ approaches a similar form ( $\sin{(x)}\  D/x
\delta$ ) but at $k_F r \ \delta/D\geq \pi$, a distance about one
thousand times greater, the Kondo length .

\begin{figure}[h]
\includegraphics[width=\columnwidth]{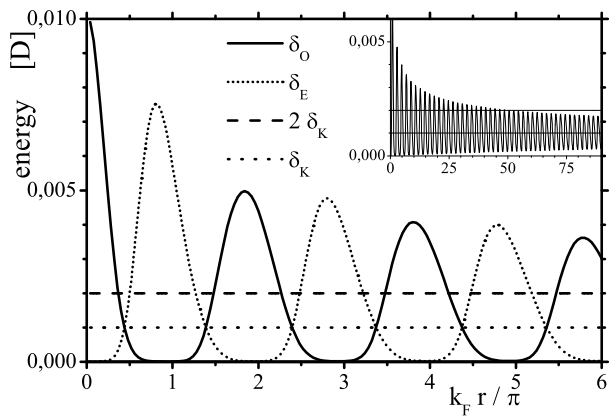}
\caption{Correlation energy gains of the odd, $\delta_O$, and even,
$\delta_E$, doublets as a function of the distance between the impurities,
for $\Sigma_R=0$ and $Jn=0.07238$. The inset
is $\delta_E$ up to very long distances, and it
periodically beats $2 \delta_K$ up to $k_F r \simeq 50 \pi$.} \label{fig2}
\end{figure}

As a function of $I_K$ and $C_D$  Eq.(\ref{es02}) becomes
\begin{equation}\label{es03}
1/Jn=[2-C_D(\delta_E, k_F r)]\ I_K(\delta_E )\ ,
\end{equation}
from which $\delta_E$ can be obtained numerically. It is interesting
to write it the Kondo way (Eq.(\ref{ekondo})),
\begin{equation}\label{es04}
\delta_E=D\ \exp{\left[ \frac{-1}{[2 - C_D(\delta_E, k_F r)]\ Jn}
\right]}\ ,
\end{equation}
which can be taken as an approximate solution by evaluating $C_D$
at $\delta_K$, or used iteratively to find the exact result with
fast convergency. Note that $C_D$ couples to the
``connectivity" factor (the $2$ in the exponent of
Eq.(\ref{ekondo})), and small changes in this factor produce huge
changes in $\delta_E$. For $\delta_O$ the sign in front of $C_D$
is positive. These energies are plotted in Fig.\ref{fig2} as a
function of $k_F r$ for $\delta_K=0.001\ D$. They alternatively
surpass $2 \ \delta_K$ up to $k_F r \simeq 50 \pi$.

This effect, $\delta_E$,$\delta_O \geq 2 \delta_K$, is present for
values of $Jn$ lower than $Jn_c=0.240$ (which gives
$\delta_K=D/8$); the lower $Jn$, the greater this effect. The
periodicity with which $|D_O\rangle$,  $|D_E\rangle$ alternate  is approximately twice the RKKY period (they are
ruled by $C_D$), and they stand in this situation up to very long
distances. These properties are due to the fact that the Doublet
 Coherence integral involves just one hole excitation. At the
intersection points one has $\delta_E,\delta_O = \delta_K$ and
$C_D = 0$ \ .

The situation for $\Sigma_R \neq 0$ can be analyzed with the same
concepts already introduced. We analyze here the even doublet
energy $E_E$ when $\Sigma_R > 0$. The analysis for $E_O$ and/or
$\Sigma_R < 0$ is very similar.

For $\Sigma_R > 0$ the lower energy configurations are the ones of
the ferro triplet, which makes the pole of $\alpha^F_q$ the
relevant one in the equation for $E_E$, Eq.(\ref{enede}). Thus,
proposing $E_E=-2 E_d - \Sigma_R-\delta_E$, Eq.(\ref{enede}) is
transformed to
\begin{eqnarray}\label{enede1}
E_d+\Sigma_R+\delta_E=\frac{3}{2}\  \textbf{v}^2 \sum_{q} \frac{1-\cos{q r}}
{\delta_E+e_q}\nonumber \\
+\frac{1}{2}\ \textbf{v}^2 \sum_{q} \frac{1+\cos{q r}}{2 \Sigma_R+\delta_E+e_q}\ ,
\end{eqnarray}
and then, using the definitions for $I_K$ and $I_X$, we have,
\begin{eqnarray}\label{enede2}
\frac{E_d+\Sigma_R+\delta_E}{n_o V^2} = \frac{3}{2}\  [I_K(\delta_E)-I_X(\delta_E)]\nonumber \\
+\frac{1}{2}\ [I_K(2 \Sigma_R + \delta_E)+I_X(2 \Sigma_R + \delta_E)]\ ,
\end{eqnarray}
a self-consistent equation for $\delta_E$ (in Eq.(\ref{enede2})
$I_X$ is also a function of $k_F r$, but only the ``energy"
dependence is quoted). If $E_d \gg \Sigma_R, \delta_E$ the left
side of Eq.(\ref{enede2}) is just $1/Jn$; if not, a new parameter
must be introduced: the $E_d/D$ ratio. In the present work we keep
it simple (just $Jn$). For $\delta_O$ (and $\Sigma_R>0$) the sign
of the $I_X$ terms are inverted. Therefore when $\delta_O$
decreases $\delta_E$ increases and viceversa, following the
oscillations of $C_D$ as a function of $k_F r$, just as in the
$\Sigma_R = 0$ case. For $\Sigma_R<0$ the dominant pole is that of
the antiferro component of the doublets and $\delta_E$, $\delta_O$
swap equations, with $|\Sigma_R|$ instead of $\Sigma_R$ and one
important difference: the bare connectivity of the antiferro
component is $1/2$, instead of the $3/2$ of the ferro one, and
hence these energy gains are lower than in the $\Sigma_R>0$ case.
\begin{figure}[h]
\includegraphics[width=\columnwidth]{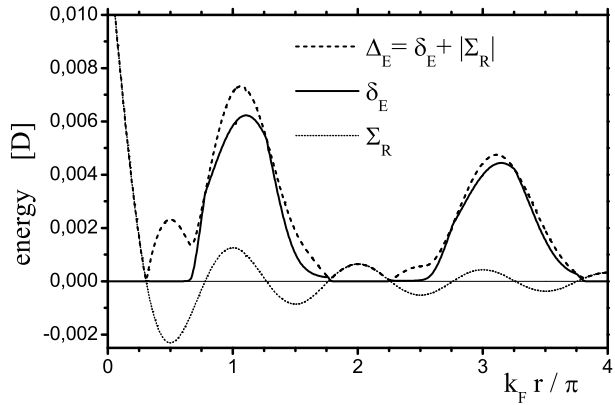}
\caption{Total energy gain of the even doublet,
$\Delta_E=|\Sigma_R|+\delta_E$, as a function of $r$. Note that in
the relevant ranges for this state the main contribution to $\Delta_E$ comes from $\delta_E$ . } \label{fig3}
\end{figure}

In Fig.\ref{fig3} we show $\delta_E$, $\Sigma_R$, and the total
energy gain of the even doublet, $\Delta_E=|\Sigma_R|+\delta_E$.
Note the kinks in $\delta_E$ at the zeros of $\Sigma_R$, those
kinks reflect the change of dominant component in $|D_E\rangle$
when $\Sigma_R$ changes sign. Nevertheless, the total energy gain 
is smooth at those points. In
the regions in which $\delta_E$ is equal to zero the doublet
``collapses": i.e. when $\delta_E=0$ the dominant $\alpha_q^{F(A)}$ diverges for $|q|=k_F$, (Eqs.(\ref{weven},\ref{alfad}), $e_{k_F}=0$)
 and thus this is the only  surviving
configuration of the ones that compose the doublet, and a FF (AF)
state with just one hole at the Fermi level is simply a FF (AF)
state. The kinks of $\Delta_E$ in these regions are because
$\Delta_E=|\Sigma_R|$. In the relevant regions ($\delta_E>\delta_O$)
usually $\delta_E>|\Sigma_R|$, thus the dominant pole
configurations (FF (AF) based) have just a little more weight in
the doublet than the ``dominated" one (AF (FF) based), and the
$\Sigma_R=0$ regime is soon reached.

\begin{figure}[h]
\includegraphics[width=\columnwidth]{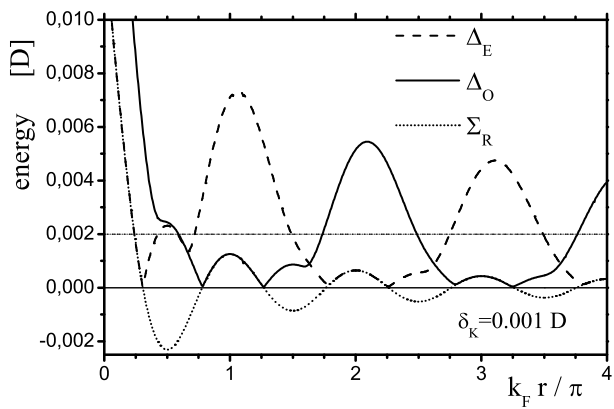}
\caption{$\Delta_E$ and $\Delta_O$ as a function of $r$, for
$\delta_K=0.001 D$. } \label{fig4}
\end{figure}

Finally, in Fig.\ref{fig4}, we plot both $\Delta_E$ and $\Delta_O$
as function of $r$, $\Sigma_R$ is also
shown ($\delta_K=0.001 D$). For this value of $Jn$, in the measure that $r$ increases the
system alternates between the odd and even doublets, just as in the $\Sigma_R=0$ case, except for a small
region above $k_F r=\pi /2$. At this point $\Sigma_R$ is near an
AF maximum and $C_D$ is at its first zero. Hence there is not
connectivity enhancement, neither for the odd, nor for the even
doublet, and they collapse into the AF state. 

In conclusion, we give a clear description of the behavior of these doublet states of the Two Anderson Impurity System and their interplay with the RKKY interaction. A remarkable point of
the model is the long distances at which the impurities remain in
a correlated state. A full understanding of this property, which
our model provides in detail, allows these systems to be used in
the design of quantum devices.

\bigskip
I am grateful to Blas Alascio for insightful inquires. I am a
fellow of the CONICET (Argentina), which partially financed this
research under grant PIP 02753/00.

\bigskip
\appendix*

\noindent \textbf{Appendix:} In 1D, $\Sigma_R$, Eq.(\ref{rkky1}), half the RKKY interaction, is
\begin{eqnarray}\label{rkky2}
\Sigma_R \simeq 2 \ \frac{\textbf{v}^4}{E_d^2}
\sum_{k\ q}\frac{\cos{(q.r)}\ \cos{(k.r)}}{(e_k+e_q)}\ \nonumber \\
=(\frac{\pi}{2}\ Jn)^2\ D\ J_R(2 k_F r)\ ,
\end{eqnarray}
where the $r$ dependence is given by\cite{litvi},
\begin{equation}\label{jr}
 J_R(2 k_F r)=\frac{2}{\pi}\ [\frac{\pi}{2}-\text{Si}(2 k_F r)]\ .
\end{equation}

The 1D Doublet Coherece integral is given by
\begin{eqnarray}\label{IX}
I_X(\delta,x)=& \cos{(x+x d)} [ \text{Ci}(x+x d)-\text{Ci}(x)] + \nonumber \\
& \sin{(x+x d)}[ \text{Si}(x+x d)-\text{Si}(x)]\ ,\ \ \ \ \ \
\end{eqnarray}
where $d=\delta/D$. $\text{Ci}$ ($\text{Si}$) is the CosIntegral
(SinIntegral) function, as defined in Mathematica$^{\circledR}$.

The upper critical value of $Jn$, given that $|C_D|\leq 1$, is
defined by the equation
\begin{equation}\label{jnc}
e^{-1/(3 Jn_c)} = 2\ e^{-1/(2 Jn_c)}\ ,
\end{equation}
which gives $Jn_c=1/6 \ln{2}$.
\bibliography{kondo}

\end{document}